# Improving a Hybrid Algorithm for APCS Hierarchical Architecture Design via Hyperparameter Optimization

Ruslan Zakirzyanov
NEXT engineering, LLC
Kazan, Russia
zr@nexteng.ru

***Abstract*—Automated process control systems, widely used in industry, have a hierarchical multi-level tree-based architecture and are built using specialized hardware and software. Constructing such a control system from off-the-shelf industrial components with predetermined characteristics is a complex combinatorial optimization problem with multiple constraints. A hybrid method for constructing a tree-based structure is considered, combining a deterministic construction algorithm and an ant colony metaheuristic for device selection. The effectiveness of the metaheuristic algorithm depends on its hyperparameters, which are proposed to be optimized. This paper presents a general problem formulation, defines optimization criteria, describes the solution method, and provides several numerical examples.**

## I. INTRODUCTION

Industrial Automated Process Control Systems (APCS) are a key element of modern large-scale industrial facilities [1], where high demands are placed on safety, reliability, and product quality. Continuous-process industries, such as chemical, petrochemical, and oil and gas production, are characterized by a particularly high degree of automation. In such systems, technological processes are performed automatically, while operators mainly supervise operation, adjust control strategies, and monitor system performance.

Large APCS are implemented using specialized hardware and software built from commercially available components, including Programmable Logic Controllers (PLCs), input/output modules, and communication and network devices. Each component has predefined characteristics, such as channel capacity, communication interfaces, memory, and processing performance.

During system design, engineers must assemble an architecture from available components that satisfies technical requirements while minimizing cost. Since component properties are fixed, system performance is largely determined by the chosen architecture, which is typically hierarchical and multi-level. In geographically distributed and large-scale installations, such systems are commonly referred to as Distributed Control Systems (DCS), where structural optimization becomes especially important. APCS may be built using single-vendor solutions or by integrating equipment from multiple manufacturers, for example within open architectures such as OPAS [2]. In practice, design decisions are often based on experience and vendor guidelines, which do not always lead to optimal configurations.

At early design stages, engineers are required to estimate system cost and scale under incomplete information about the controlled process. These estimates directly affect project planning and resource allocation.

The structural synthesis of APCS can be formulated as a constrained combinatorial optimization problem. Exact methods are applicable only to relatively small instances due to their computational complexity. For large-scale systems, approximate approaches are more practical, providing near-optimal solutions within reasonable time. Among them, metaheuristic algorithms have shown strong performance [3], including both local search methods and population-based techniques inspired by evolutionary processes and swarm intelligence.

In previous studies [4-6], an Ant Colony Optimization (ACO)-based framework was proposed for constructing tree-structured APCS architectures under limited information. In this paper, that approach is extended by incorporating an adaptive heuristic mechanism and a multi-objective, Pareto-based hyperparameter tuning strategy. Numerical experiments demonstrate improved feasibility and reduced variability of the obtained solutions.

## II. RELATED WORK

Structural optimization of engineering systems remains an important research area, with structural synthesis problems studied across various domains, including distributed control system design [7].

Among approximate methods, metaheuristic algorithms have attracted significant attention. A wide range of bio-inspired techniques has been developed, such as Genetic Algorithms (GA), Artificial Bee Colony (ABC), Grey Wolf Optimizer (GWO), Ant Colony Optimization (ACO), Particle Swarm Optimization (PSO), and others [8]. An overview of their evolution is presented in [9]. These methods have been successfully applied to diverse engineering optimization problems.

In addition to standalone approaches, hybrid metaheuristics that combine different optimization strategies have been investigated, particularly combinations of swarm intelligence and local search [10]. More recent studies also explore integrating metaheuristics with machine learning techniques. Hyperparameter tuning remains a critical factor influencing algorithm performance, and methods for automated generation of problem-specific optimization strategies have also been proposed.

Ant Colony Optimization, introduced by Dorigo for the Traveling Salesman Problem [11], has developed into a flexible metaheuristic with numerous extensions [12]. It has been applied to different tasks such as controller tuning, routing, and facility location [13-15].

The APCS structural optimization problem considered in this work was previously formulated in [4, 6]. However, despite the extensive literature on metaheuristics and structural optimization, the specific problem of optimizing hierarchical APCS architectures remains insufficiently studied. Existing works either focus on general-purpose methods or consider structural synthesis in a limited form, which motivates the development of specialized metaheuristic approaches for this problem.

## III. Problem Formulation

A detailed formal formulation of the optimal APCS architecture synthesis problem is presented in [5]. The main components of the model are outlined below.

The APCS architecture is represented as a tree (i.e., an acyclic graph) $G = (\mathcal{V}, \mathcal{E})$, where $\mathcal{V}$ is the set of nodes corresponding to devices $v \in \mathcal{V}$, and $\mathcal{E}$ is the set of edges representing communication links between them. An example of such an architecture is shown in Fig. 1.

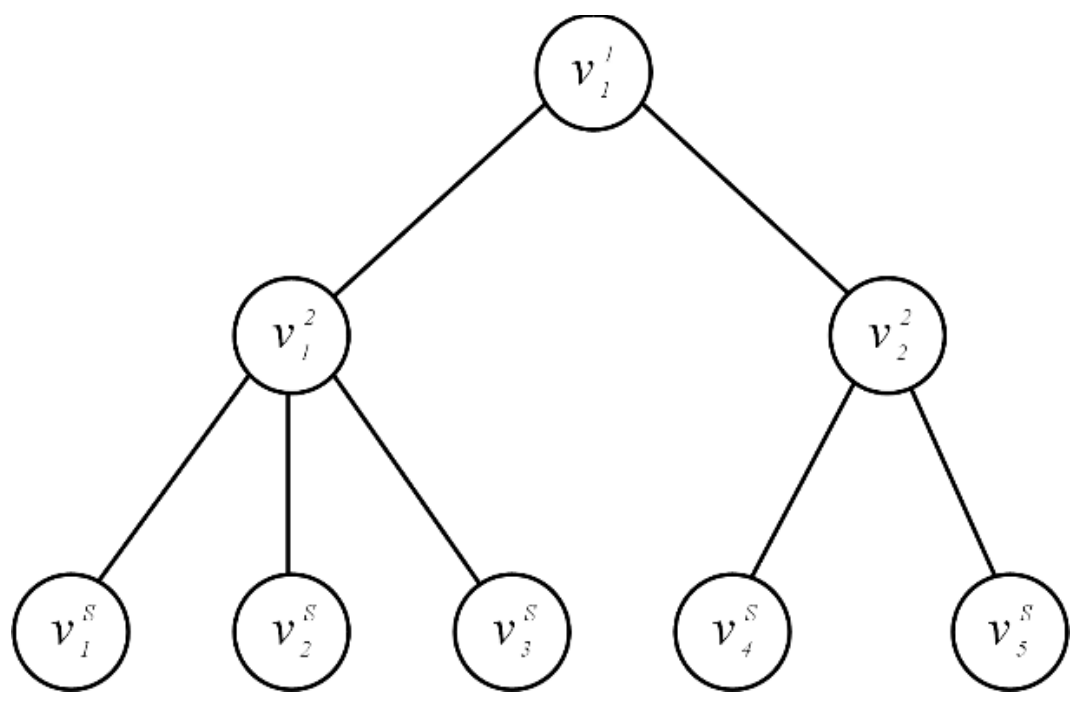

Fig. 1. Hierarchical architecture of the APCS

The number of hierarchical levels is specified a priori by the system designer.

Let $\mathcal{U} = \{u_1, \dots, u_U\}$ denote the set of available device types. Each device executes a cyclic program that includes three main stages: data acquisition, data processing (i.e., execution of control algorithms), and data transmission. Devices are categorized into two classes: processors and repeaters.

Each device type $u_i \in \mathcal{U}$ is characterized by the following parameter vector:

$$u_i = (C_i, N_i, R_i, P_i, T_i, y_i, M_i, \tau_i), \quad i = 1, \dots, U, \quad (1)$$

where $C_i \in \mathbb{R}^+$ denotes the device cost, $N_i \in \mathbb{N}$ – the number of physical channels, $R_i \in \mathbb{R}^+$ – the available memory, $P_i \in [0,1]$ – the failure probability, $T_i \in \mathbb{R}^+$ – the instruction execution time, $y_i \in \{0,1\}$ – the device role (0 corresponds to a repeater, 1 to a processor), $M_i \in \mathbb{N}$ – the maximum number of child nodes, and $\tau_i \in \mathbb{R}^+$ – the communication delay (relevant for repeaters).

Let $\mathcal{A} = \{a_1, \dots, a_A\}$ denote the set of control loops. Each loop is defined as:

$$a_j = (n_j, r_j, w_j), \quad j = 1, \dots, A, \quad (2)$$

where $n_j \in \mathbb{N}$ is the number of physical signals, $r_j \in \mathbb{R}^+$ – the required memory, and $w_j \in \mathbb{N}$ – the computational load expressed in the number of program instructions.

To describe loop allocation, binary decision variables $x_{va} \in \{0,1\}$ and $z_{va} \in \{0,1\}$ are introduced. The variable $x_{va} = 1$ indicates that loop $a$ a is physically connected to leaf node $v$, and 0 otherwise. The variable $z_{va} = 1$ indicates that loop $a$ is processed at node $v$, and 0 otherwise.

The formulation includes constraints related to device capacity, memory limitations, execution time, system reliability, and consistency of the hierarchical structure.

The objective is to determine a feasible hierarchical architecture $G$ that minimizes the total system cost:

$$C^* = \min_{G \in \mathcal{G}} \sum_{v \in \mathcal{V}(G)} C_v, \quad (3)$$

subject to the imposed constraints. Here, $\mathcal{G}$ denotes the set of all admissible hierarchical graphs that can be constructed from the available device types.

## IV. Proposed Method

The APCS architecture synthesis problem is formulated as a constrained combinatorial optimization task, where direct use of population-based metaheuristics (e.g., GA, PSO, ABC) typically produces many infeasible solutions due to constraint violations. This leads to inefficient search and increased computational cost. To mitigate this issue, a hybrid approach is proposed, integrating a deterministic tree construction procedure with a metaheuristic-based device selection mechanism.

The deterministic algorithm incrementally builds a tree-structured architecture. First, the backbone (stem) of the hierarchy is formed. Then, control loops are assigned to leaf nodes. If capacity constraints are violated, additional branches and devices are introduced as needed. Since feasibility is enforced during construction, invalid architectures are largely avoided. At specific decision points, where a device type must be chosen, selection is performed using Ant Colony Optimization.

ACO is well suited for such constructive problems. At each decision step, a candidate device $i$ is selected probabilistically:

$$P_i = \frac{\tau_i^{\alpha} \eta_i^{\beta}}{\sum_{k=0}^{N} \tau_k^{\alpha} \eta_k^{\beta}}, \quad (4)$$

where $\tau_i$ denotes pheromone intensity, $\eta_i$ – heuristic desirability, and $\alpha$, $\beta$ are weighting parameters. Pheromone values are updated iteratively:

$$\tau_i \leftarrow (1 - \rho)\tau_i + \Delta\tau_i, \quad (5)$$

where $\rho \in (0,1)$ controls evaporation.

Compared to random greedy and Monte Carlo strategies, ACO demonstrates higher solution quality but still struggles near feasibility boundaries, where valid configurations become scarce. This motivates the use of an adaptive heuristic.

A basic cost-driven heuristic (6) tends to favor inexpensive devices but may lead to infeasible structures due to insufficient capacity.

$$\eta_i = \frac{1}{c_i}, \quad (6)$$

To address this, a level-dependent heuristic is introduced. For leaf nodes ($s = S$), priority is given to channel capacity:

$$\eta_i = \frac{N_i^2}{c_i}, \quad s = S. \quad (7)$$

For internal levels ( $s < S$ ), branching capability is additionally considered (8).

$$\eta_i = \frac{N_i^2 M_i}{c_i}, \quad s < S. \quad (8)$$

The heuristic is further adapted based on the number of unassigned loops and remaining expansion capacity. If the number of required assignments exceeds available branching resources, candidates with limited connectivity are penalized to avoid premature saturation of the structure.

The performance of ACO is sensitive to its parameters, particularly $\alpha$, $\beta$, and $\rho$, which influence both feasibility and solution stability. Therefore, parameter tuning is formulated as a multi-objective optimization problem and solved using a Pareto-based approach, enabling a trade-off between feasibility and robustness.

Finally, a local search (LS) refinement is applied to the best solution. This involves replacing a randomly selected device with an alternative type; the modification is accepted if it improves the objective while maintaining feasibility.

## V. Discussion

The proposed framework was implemented in Python with a Qt-based graphical interface, allowing flexible configuration of device types, control loop parameters, hierarchical depth, and ACO settings (number of iterations, ants, and parameters $\alpha, \beta, \rho$).

Six device types were considered (Table I), including two processor units analogous to industrial PLCs and four repeater-type devices representing I/O modules with varying channel capacities. Processor units are relatively costly and do not provide direct physical I/O, whereas repeater devices support between one and six signals and include communication interfaces.

TABLE I. Device Types

| Dev. Type | Device Parameters | | | | | | | |
|---|---|---|---|---|---|---|---|---|
| | $C_i$ | $N_i$ | $R_i$ | $P_i$ | $T_i$ (s) | $y_i$ | $M_i$ | $\tau_i$ (s) |
| $u_1$ | 1300 | 0 | 512 | 0.001 | 0.007 | 1 | 4 | 0 |
| $u_2$ | 950 | 0 | 256 | 0.003 | 0.003 | 1 | 3 | 0 |
| $u_3$ | 110 | 6 | - | 0.0015 | - | 0 | 6 | 0.02 |
| $u_4$ | 90 | 4 | - | 0.002 | - | 0 | 8 | 0.03 |
| $u_5$ | 70 | 1 | - | 0.002 | - | 0 | 8 | 0.02 |
| $u_6$ | 55 | 2 | - | 0.002 | - | 0 | 4 | 0.04 |

All experiments were conducted on a workstation with an Intel Core i5 CPU, 16 GB RAM, and SSD storage. An example of the resulting optimal architecture for $A = 100$ control loops and $S = 3$ hierarchy levels is shown in Fig. 2.

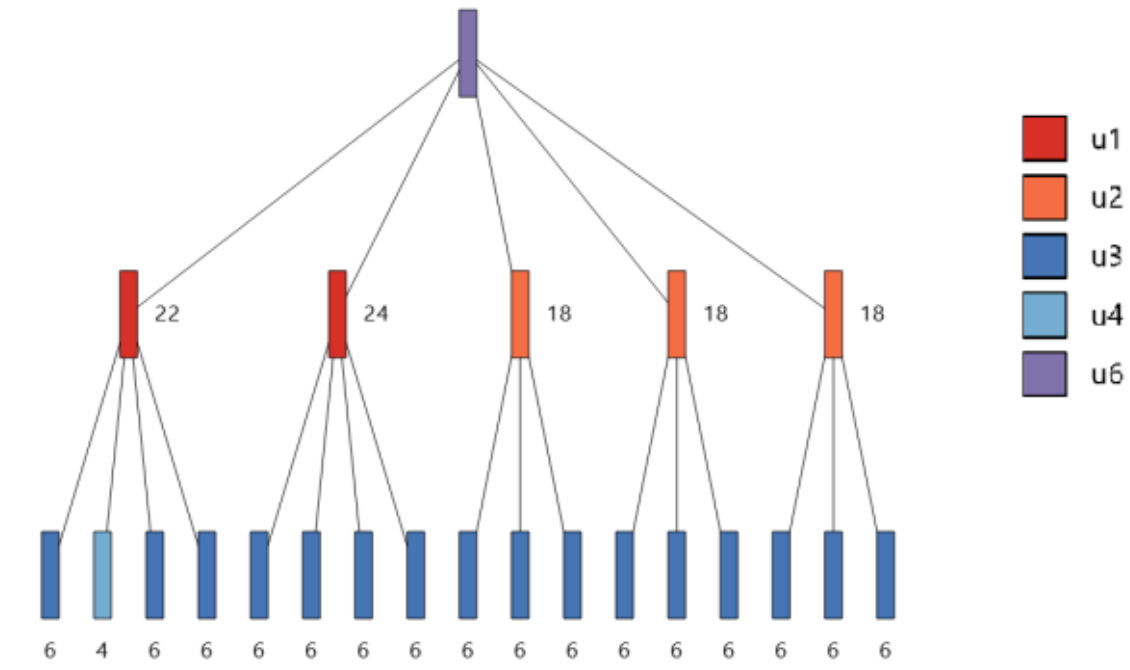

Fig. 2. APCS architecture for $A = 100$ and $S = 3$

Several ACO variants were evaluated, including Ant System (AS), Min-Max Ant System (MMAS), and Elite Ant System (EAS), each tested with both the basic cost-driven heuristic and the proposed adaptive heuristic. The adaptive versions are denoted as "new".

Algorithm performance was assessed using feasibility $W$ (percentage of successful runs) and relative variability $CV^*$, defined as the coefficient of variation normalized by the best solution. Since ACO performance strongly depends on parameters, a multi-objective tuning procedure based on NSGA-II [16] was applied. The resulting Pareto front (Fig. 3) was used to select parameter values prioritizing feasibility, yielding $\alpha = 2.2546$, $\beta = 8.3215$, and $\rho = 0.7856$. These configurations are labeled as "Optimized".

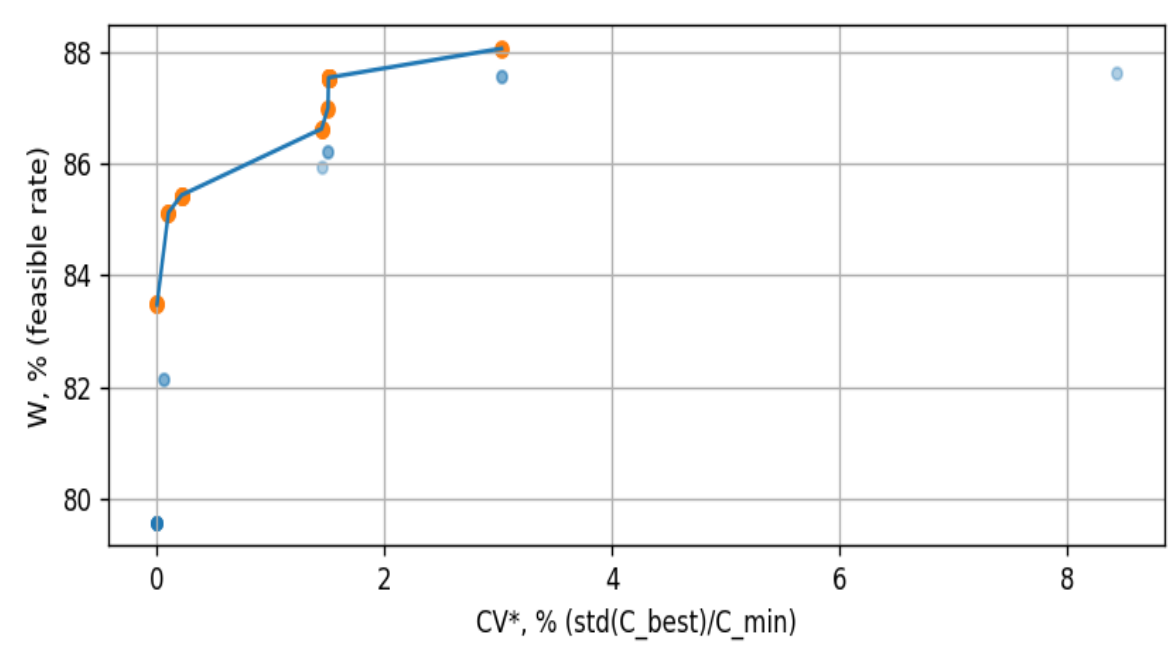

Fig. 3. Pareto front for ACO parameter tuning

TABLE II. Experimental Results

| Exp. No | Strategy | $C_{min}$ | $C_{avg}$ | $CV^*(\%)$ |
|---|---|---|---|---|
| 1 | AS | 7355 | 7411 | 1.62 |
| 2 | MMAS | 7355 | 7370 | 0.25 |
| 3 | Elite MMAS | 7355 | 7361 | 0.15 |
| 4 | Elite MMAS new | 7355 | 7368 | 0.24 |
| 5 | AS new Optimized | 8090 | 8274 | 3.05 |
| 6 | AS new Optimized + LS | 7740 | 7880 | 2.84 |
| 7 | Elite MMAS new Optimized | 8090 | 8141 | 1.34 |
| 8 | Elite MMAS new Optimized + LS | 7355 | 7384 | 1.05 |

The results show that only optimized variants maintain high feasibility across the full range of control loop counts. However, increasing the influence of heuristic or pheromone components may reduce solution stability and cause premature convergence.

To address this issue, a local search (LS) stage was incorporated. The LS performs a simple neighborhood move by replacing a randomly selected node with an alternative device type, accepting the change if it improves cost while preserving feasibility. Experimental results are summarized in Table II.

Convergence curves (Fig. 4, 5) illustrate the evolution of solution cost over iterations, showing both average and best performance across multiple runs. The results indicate that the optimized ACO combined with LS achieves consistently high feasibility, low variability, and competitive cost values across the considered scenarios.

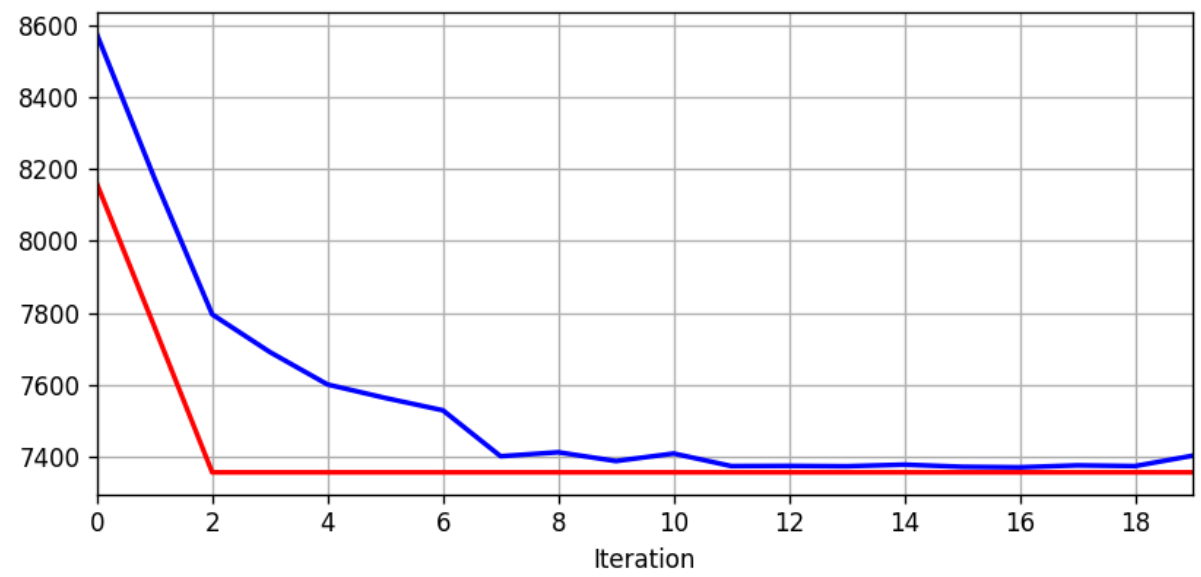

Fig. 4. Convergence of the Elite MMAS (adaptive heuristic) algorithm

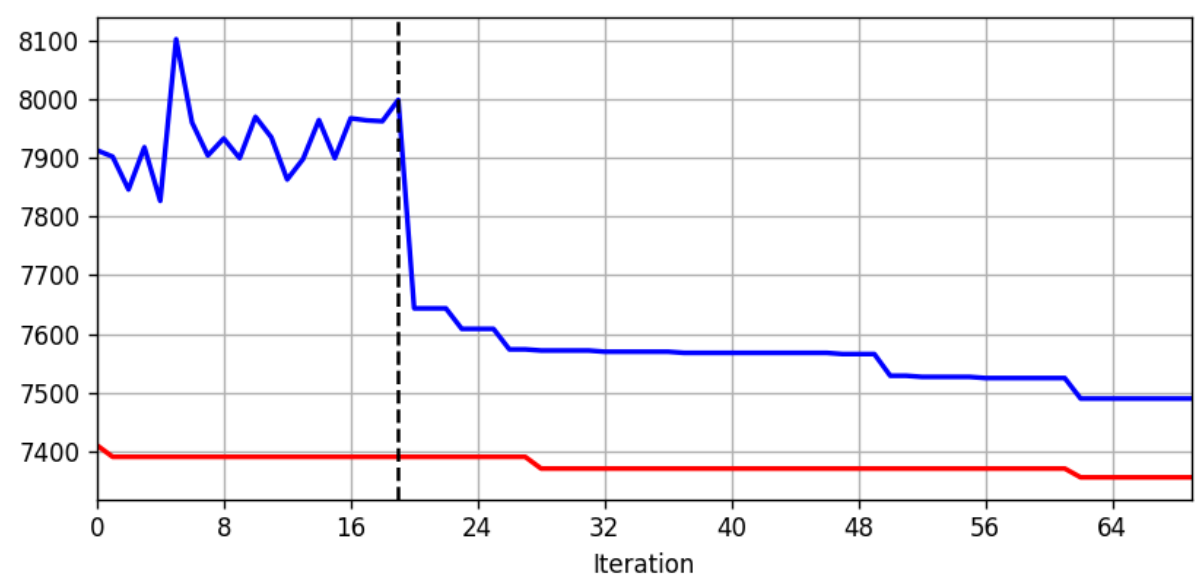

Fig. 5. Convergence of the optimized Elite MMAS with local search

## VI. Conclusion and Future Work

The design of an optimal APCS architecture under limited initial information can be formulated as a constrained combinatorial optimization problem. Due to its high computational complexity, metaheuristic approaches are well suited for this task. To guarantee feasibility under multiple structural constraints, a deterministic construction procedure was integrated with ACO for device selection. To enhance performance in regions close to feasibility boundaries, an adaptive heuristic was introduced, and key ACO parameters were tuned using a multi-objective strategy that balances feasibility and solution stability. In addition, a local search stage was incorporated to reduce the risk of premature convergence.

Computational experiments show that the proposed hybrid framework increases the proportion of feasible solutions over the admissible range of control loop counts while maintaining competitive cost and low variability. The developed approach also enables explicit modeling of communication channels within the APCS architecture, allowing the system to be analyzed within the cyber-physical systems paradigm.

Future work will focus on improving the fidelity of the system model to better capture real device characteristics, evaluating alternative optimization techniques, and exploring integration with machine learning methods.